\documentclass[8.5pt,twoside,twocolumn]{article}

\usepackage[super,sort&compress,comma]{natbib}
\usepackage{times,mathptm}
\usepackage{sectsty}
\usepackage{balance}
\usepackage{graphicx}
\usepackage{lastpage}
\usepackage[format=plain,justification=raggedright,singlelinecheck=false,font=small,labelfont=bf,labelsep=space]{caption}
\usepackage{fancyhdr}
\usepackage{amsmath}
\usepackage{amssymb}
\usepackage{amsthm}
\usepackage{bm}

\begin{document}

\thispagestyle{plain}
\fancypagestyle{plain}{
\renewcommand{\headrulewidth}{1pt}}
\renewcommand{\thefootnote}{\fnsymbol{footnote}}
\renewcommand\footnoterule{\vspace*{1pt}%
\hrule width 3.4in height 0.4pt \vspace*{5pt}}
\setcounter{secnumdepth}{5}

\makeatletter
\def\subsubsection{\@startsection{subsubsection}{3}{10pt}{-1.25ex plus -1ex minus -.1ex}{0ex plus 0ex}{\normalsize\bf}}
\def\paragraph{\@startsection{paragraph}{4}{10pt}{-1.25ex plus -1ex minus -.1ex}{0ex plus 0ex}{\normalsize\emph}}
\renewcommand\@biblabel[1]{#1}
\renewcommand\@makefntext[1]%
{\noindent\makebox[0pt][r]{\@thefnmark\,}#1}
\makeatother
\renewcommand{\figurename}{\small{Fig.}~}
\sectionfont{\large}
\subsectionfont{\normalsize}

\fancyfoot{}
\fancyfoot[RO]{\footnotesize{\sffamily{1--\pageref{LastPage} ~\textbar  \hspace{2pt}\thepage}}}
\fancyfoot[LE]{\footnotesize{\sffamily{\thepage~\textbar\hspace{3.45cm} 1--\pageref{LastPage}}}}
\fancyhead{}
\renewcommand{\headrulewidth}{1pt}
\renewcommand{\footrulewidth}{1pt}
\setlength{\arrayrulewidth}{1pt}
\setlength{\columnsep}{6.5mm}
\setlength\bibsep{1pt}

\twocolumn[
  \begin{@twocolumnfalse}
\noindent\LARGE{\textbf{A chiral rhenium complex with predicted high parity violation effects:
synthesis, stereochemical characterization by VCD spectroscopy and quantum chemical calculations}}
\vspace{0.6cm}

\large
\noindent \textbf{Nidal~Saleh},\emph{$^{a}$}
          \textbf{Samia~Zrig},\emph{$^{a,b}$}
          \textbf{Thierry~Roisnel},\emph{$^{a}$}
          \textbf{Laure~Guy},\emph{$^{c}$}
          \textbf{Radovan~Bast},\emph{$^{d}$}
          \textbf{Trond~Saue},\emph{$^{d}$} \\
          \textbf{Beno\^it~Darqui\'e},\emph{$^{e}$}
          \textbf{and}
          \textbf{Jeanne~Crassous}$^{\ast}$\emph{$^{a}$}
\normalsize
\vspace{0.5cm}

\vspace{0.6cm}

\noindent \normalsize{
With their rich electronic, vibrational, rotational and hyperfine
structure, molecular systems have the potential to play a decisive
role in precision tests of fundamental physics. For example, electroweak
nuclear interactions should cause small energy differences between
the two enantiomers of chiral molecules, a signature of
parity symmetry breaking. Enantioenriched oxorhenium(VII) complexes
\emph{S}-(-)- and \emph{R}-(+)-\textbf{3} bearing a chiral 2-methyl-1-thio-propanol
ligand have been prepared as potential candidates for probing molecular
parity violation effects via high resolution laser spectroscopy of
the Re=O stretching. Although the rhenium atom is not a stereogenic
centre in itself, experimental vibrational circular dichroism (VCD)
spectra revealed a surrounding chiral environment, evidenced by the
Re=O bond stretching mode signal. The calculated VCD spectrum of the \emph{R} enantiomer
confirmed the position of the sulfur atom \emph{cis} to the methyl, as
observed in the solid-state X-ray crystallographic structure, and showed
the presence of two conformers of comparable stability. Relativistic
quantum chemistry calculations indicate that the vibrational shift
between enantiomers due to parity violation is above the target sensitivity
of an ultra-high resolution infrared spectroscopy experiment under
active preparation.
}
\vspace{0.5cm}
 \end{@twocolumnfalse}
  ]

\footnotetext{\emph{$^{a}$~ Institut des Sciences Chimiques de Rennes, UMR 6226 CNRS - Universit{\'e} de Rennes 1, Campus de Beaulieu, F-35042 Rennes Cedex, France. Fax: (+33) 2-23-23-69-39; Tel: (+33) 2-23-23-57-09; E-mail: jeanne.crassous@univ-rennes1.fr}}
\footnotetext{\emph{$^{b}$~ Present address: ITODYS, UMR 7086 CNRS, Universit{\'e} Paris Diderot, Sorbonne Paris Cit{\'e}, F-75205 Paris Cedex 13, France}}
\footnotetext{\emph{$^{c}$~ Laboratoire de Chimie, UMR 5182 Ecole Normale Sup{\'e}rieure de Lyon-CNRS,
F-69364 Lyon 07, France}}
\footnotetext{\emph{$^{d}$~ Laboratoire de Chimie et Physique Quantiques, UMR 5626, CNRS et Universit{\'e} de Toulouse 3 (Paul Sabatier), 118 route de Narbonne, F-31062 Toulouse, France}}
\footnotetext{\emph{$^{e}$~ Laboratoire de Physique des Lasers, Universit{\'e} Paris 13, Sorbonne Paris Cit{\'e}, CNRS, F-93430, Villetaneuse, France}}

\section{Introduction}\label{sec:intro}

Chirality is a challenge in diverse domains of chemistry, such as in the
development and synthesis of molecules for pharmaceuticals,
\cite{Francotte2006} agrochemicals, flavors, in nanotechnology
\cite{Amabilino2009,Amabilino2009b} or in asymmetric catalysis.
\cite{Mikami2007} Furthermore, studying unconventional chiral systems may be of great help for understanding the origin of prebiotic homochirality.\cite{Viedma2011} More recently, very simple chiral molecules have been the
object of interest for parity
violation (PV) measurements, aiming at probing a tiny energy difference
between the right- and left-handed enantiomers of
a chiral molecule.\cite{Wagniere2007,Quack2002,Avalos2000} During the last
decade, we have been interested in observing this effect using an ultra-high
resolution infra-red (IR) spectroscopy technique on a supersonic molecular beam.
\cite{Crassous2003,Crassous2005,Chardonnet2006,Darquie2010,Stoeffler2011}
This collaboration between theoretical chemists,
synthetic chemists, and spectroscopists, aiming at observing a vibrational PV
frequency shift between enantiomers of chiral molecules, has recently been
reviewed.\cite{Darquie2010}

\begin{figure}[h]
    \centering
    \includegraphics[width=\columnwidth]{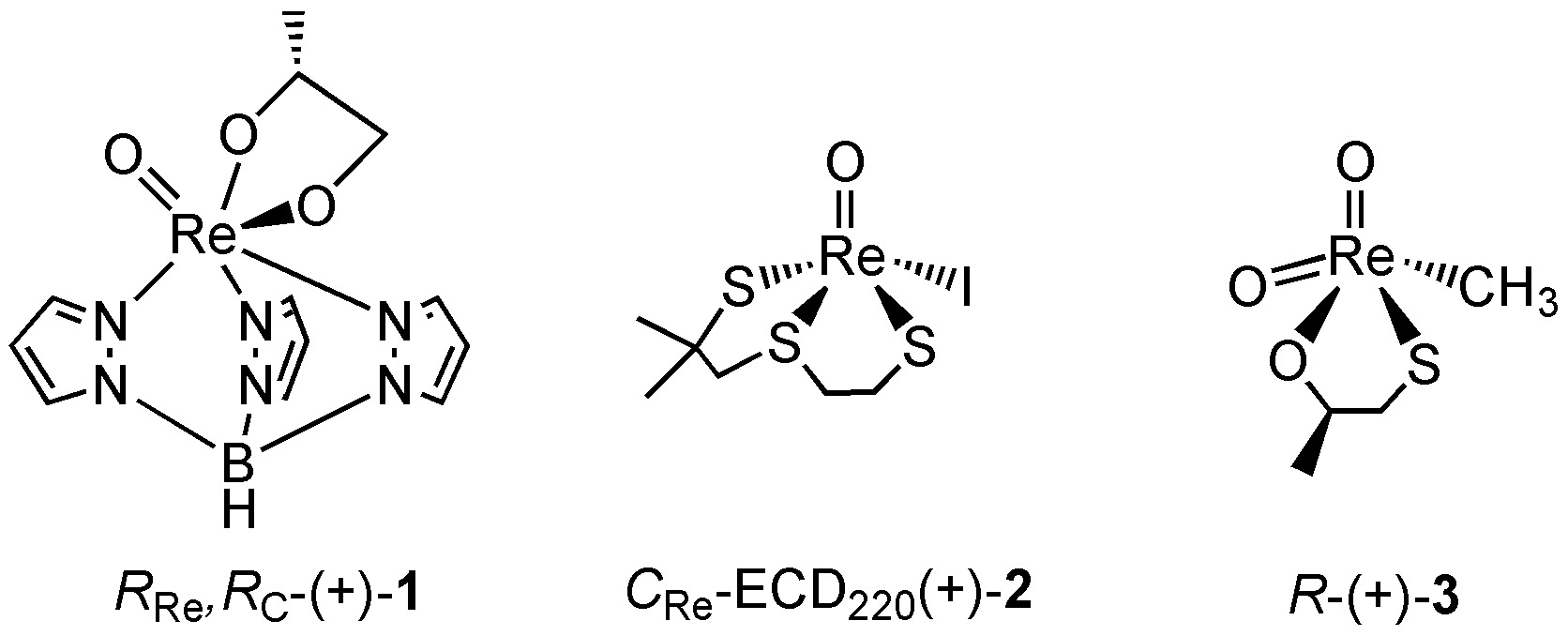}
    \caption{
             \label{fig:oxorhenium_complexes}
             Examples of chiral oxorhenium complexes studied for parity violation effects
             (refs.~\citenum{Lassen2006,DeMontigny2009,DeMontigny2010} and this work).
            }
\end{figure}

A number of chiral compounds with heavy metals as the chiral centers have been
theoretically studied for parity violation.
\cite{Laerdahl2000,Schwerdtfeger2004,Schwerdtfeger2003,Faglioni2003,Bast2003,Figgen2010}
In this context ``chiral at metal'' oxorhenium complexes have proven
particularly suitable heavy transition metal complexes because they may fulfill
all the requirements needed for a successful PV experiment. Indeed, the best
candidate for our experimental test should i) show a large PV vibrational
frequency difference of an intense fundamental band preferably within the
CO$_2$ laser operating range (900--1100 cm$^{-1}$), ii) be available in large
enantiomeric excess or, ideally, in enantiopure form and at the gram-scale,
iii) have a reasonably simple structure so as to facilitate the spectroscopy and maintain a favourable partition function, and iv) allow the production of the supersonic
expansion, thus sublime without decomposition.\cite{Darquie2010} Only a few
simple ``chiral at metal'' rhenium complexes are known in the literature.
\cite{Bock2006,Merrifield1982,Buhro1983,Lassen2006,DeMontigny2009,DeMontigny2010,Faller2000,Tazacs2012,Rybak2003,Jain2008}
Two classes of chiral complexes have been studied for PV experiments (see Figure \ref{fig:oxorhenium_complexes}): i) oxorhenium complexes such as \textbf{1}, bearing hydrotris(1-pyrazolyl)borate
(Tp) ligand and a chiral bidentate ligand,\cite{Lassen2006} and ii) enantiopure
``3+1'' oxorhenium complexes bearing a tridentate sulfurated ligand and a
monodentate halogen (such as \textbf{2}) or a chalcogenated ligand.
\cite{DeMontigny2009,DeMontigny2010} Methyltrioxorhenium (MTO) is an efficient
catalyst for reactions such as for instance epoxidation of olefins.
\cite{Owen2000} More interestingly for our purposes, it sublimes very easily\cite{Stoeffler2011}
(vapour pressure of a few tenths of mbars at room temperature). Furthermore, a
few examples of sublimable MTO derivatives have been described in the
literature \cite{Jain2008} and the strategy of preparing simple chiral easily
sublimable MTO derivatives seems an appealing route to find a suitable candidate
molecule for the experimental PV test. Our target molecule was therefore the
enantiopure oxorhenium complex \textbf{3} which is a chiral derivative of
MTO bearing a 2-methyl-thiopropanolate ligand.  In this
article we detail the synthesis, the stereochemical characterization and vibrational circular dichroism (VCD)
spectroscopy of complex \textbf{3} to examine the dissymmetric environment around
the rhenium atom. We furthermore carry out relativistic quantum chemical calculations of
the vibrational frequency shift associated with the parity violation energy
difference between enantiomers of complex \textbf{3}. Experimental and computational details are given
in sections \ref{sec:exp} and \ref{sec:comp}, respectively.

\section{Synthesis}\label{sec:synthesis}

The synthesis of the enantiomers of the 2-methyl-thiopropanol (\emph{R}-(--)-
and \emph{S}-(+)-\textbf{4}) ligands was achieved using enantiomerically pure
propylene oxide as the starting precursor as described in Scheme 1. The
regioselective opening of \emph{R}-(+)- and \emph{S}-(--)-propylene oxide
with trityl thiol \cite{Harding2000} gave respectively the alcohols \emph{R}-(+)-
and \emph{S}-(--)-\textbf{5} with 90\% yield. The subsequent deprotection followed
by \emph{in situ} oxidation with iodine yielded quantitatively disulfide
\emph{R,R}-(--)- and \emph{S,S}-(+)-\textbf{6}, which were finally reduced
by LiAlH$_4$ to \emph{R}-(--)- and \emph{S}-(+)-\textbf{4 }with 84\% yield.
Their optical rotation and all spectroscopic characteristics are in agreement
with already published results.\cite{Fox2001} The (+)- and (--)-\textbf{3}
enantiomeric complexes were then prepared by reacting the enantiopure
\emph{R}-(+)- and \emph{S}-(--)-\textbf{4 }with MTO in dry dichloromethane
according to a previously reported procedure \cite{Dixon2002} as depicted in Scheme 2.

\begin{figure}[h]
    \centering
    \includegraphics[width=\columnwidth]{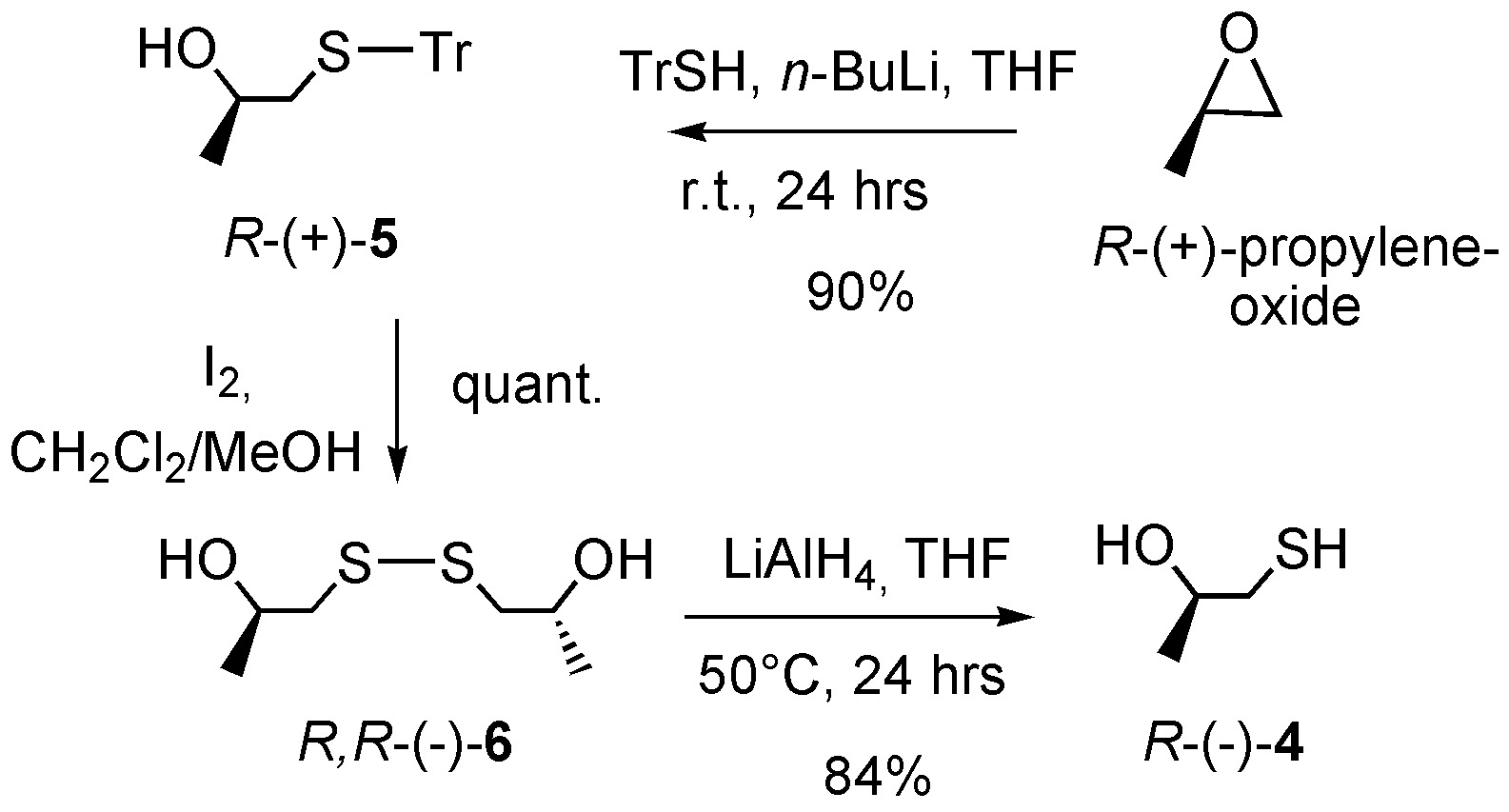}
    \caption*{
             \label{fig:scheme1}
             \textbf{Scheme 1}. Synthesis of the enantioenriched ligand 2-methyl-1-thiopropanol
             \emph{R}-(--)-\textbf{4} from \emph{R}-(+)-propylene oxide.
            }
\end{figure}

\begin{figure}[h]
    \centering
    \includegraphics[width=\columnwidth]{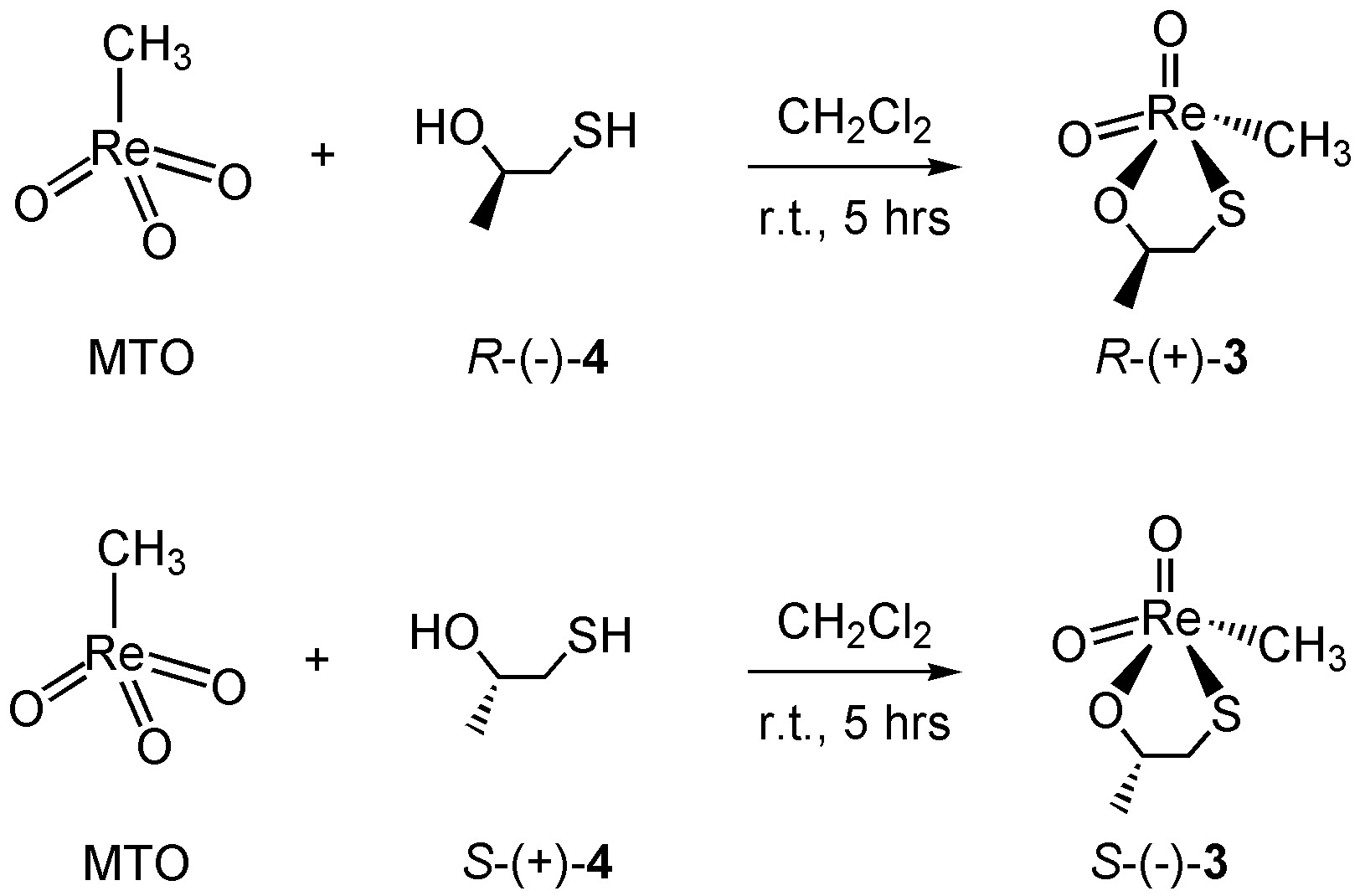}
    \caption*{
             \label{fig:scheme2}
             \textbf{Scheme 2} Synthesis of the enantioenriched oxorhenium complexes
             \emph{R-(+)-} and \emph{S}-(--)-\textbf{3} from MTO and
             enantioenriched 2-methyl-1-thiopropanol
             \emph{R}-(--)- and \emph{S}-(+)-\textbf{4}.
            }
\end{figure}

\section{X-ray crystallography}

\begin{figure}[h]
    \centering
    \includegraphics[width=\columnwidth]{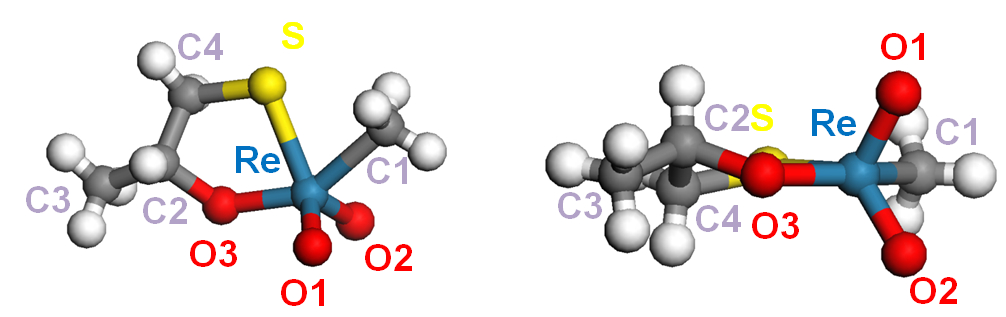}
    \caption{
             \label{fig:xray}
            X-ray crystallographic structure of the complex
             \emph{R}-(+)-\textbf{3}.
            }
\end{figure}

Single crystals of a racemic sample of the oxorhenium(VII) complex \textbf{3} were obtained after
sublimation under vacuum. Due to significant disorder of the methyl group
coming from the two possible configurations of the stereogenic carbon center,
the X-ray structure of complex \textbf{3} was solved in the non-centrosymmetric
P21 space-group. A final Flack parameter of 0.46 was obtained,
indicating the presence of a racemic mixture. As depicted in Figure \ref{fig:xray}, the rhenium center is pentacoordinated, the C1
atom is \emph{trans} to the O3 oxygen, and the Re,O3,S,C1 lie in the same
plane, while the two oxo groups are symmetrically placed on each side of this
plane (for example, angles O3--Re--O1: 100.66$^{\circ}$ and O3--Re--O2: 105.03$^{\circ}$ for one of the four independent
molecules of the unit cell). Consequently the oxygen atoms O1 and O2 only differ in their
chemical environment due to the proximity of the asymmetric carbon C2. The Re--O
bond lengths in the two oxo groups are classic (1.675 {\AA} to 1.687 {\AA}). Since oxorhenium
complexes most often crystallize in a square pyramidal geometry, with an oxygen
atom of the Re=O bond being placed in the apical position, the geometry around
the rhenium atom is rather uncommon.\cite{Espenson2002}  Finally, in the solid
state the methyl group C3 is placed in the equatorial position, and the
dihedral angle O3--C2--C4--S within the five-membered ring is about --51$^{\circ}$.
The conformational space of complex \textbf{3} has been further explored by
density functional theory (DFT) calculations, as discussed in the next section.

\section{VCD spectroscopy}\label{sec:vcd}

For our fundamental physics purposes we would like simple chiral molecules containing
one or more heavy metals, preferably as a stereogenic center or at least in its vicinity.
The former is nonetheless not straightforwardly realized.\cite{DeMontigny2009,DeMontigny2010}
From the experimental viewpoint, VCD
spectroscopy, which has been under-explored in chiral complexes, can be a very
powerful tool to study the dissymmetric environment around the metal center.
\cite{Freedman2003,Stephens2000,Polavarapu2004,Chamayou}
In order to get more insight into the latter, the IR and VCD spectra
of complex \textbf{3} were calculated and compared to
experiment.

\begin{figure}
    \includegraphics[width=\columnwidth]{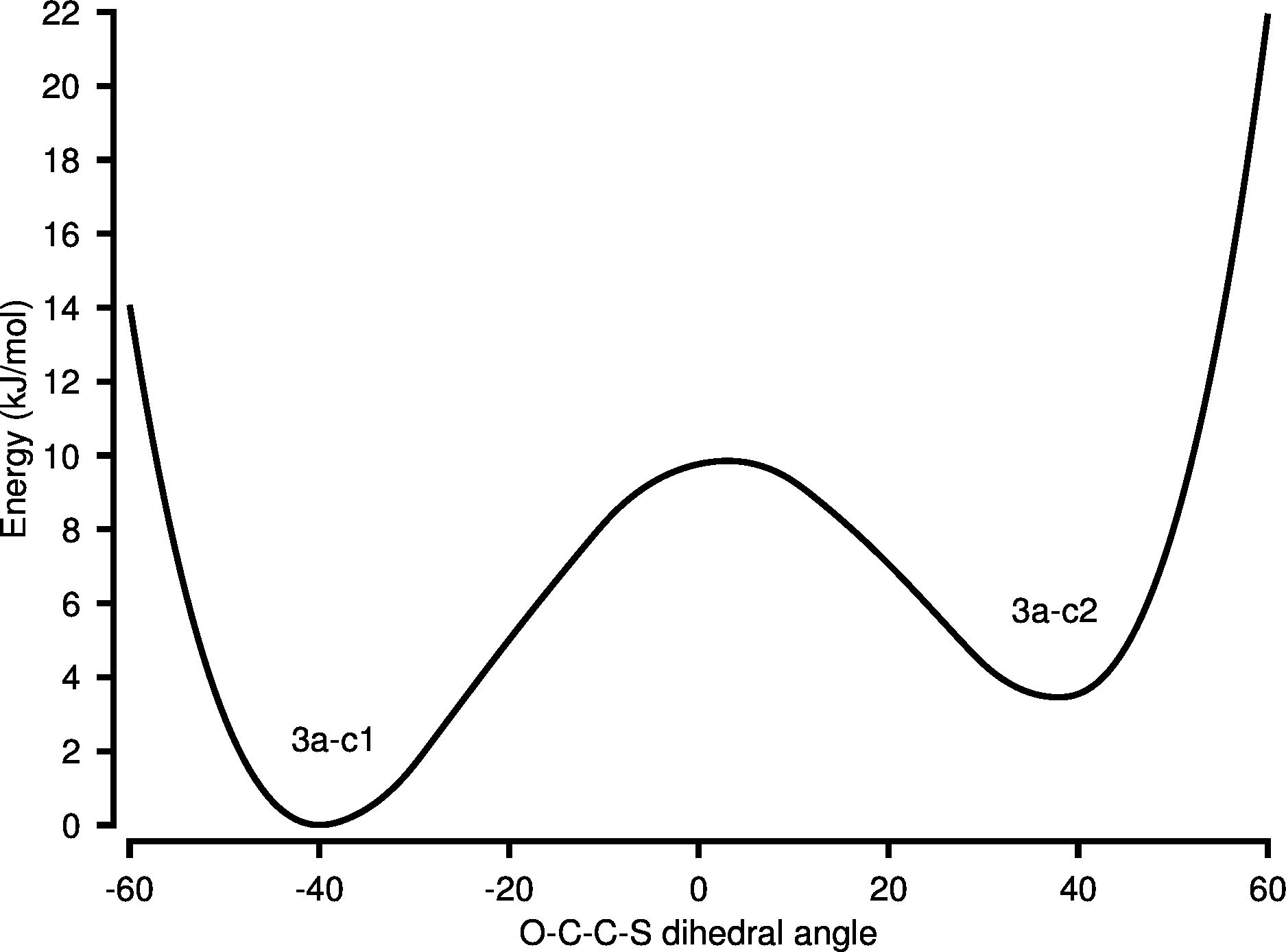}
    \caption{
             \label{fig:scan}
             Potential energy scan of complex \textbf{3} along the O3--C2--C4--S dihedral angle
             (B3LYP, def2-TZVPP basis, all other coordinates relaxed).
            }
\end{figure}

Our first computational goal, however, was to find a set of stable conformers
of complex \textbf{3} for a subsequent calculation of IR and VCD spectra.
Using the \emph{R} enantiomer and starting from the X-ray structure in which
the S--Re bond is placed in \emph{cis} position to the Re--Me one, we carried
out a potential energy scan along the O3--C2--C4--S dihedral angle at the DFT
level using the B3LYP functional (def2-TZVPP basis\cite{def2-TZVPP}).  We
identified two stable conformers, \textbf{3a-c1} and \textbf{3a-c2} at the
O3--C2--C4--S dihedral angles --39.8$^{\circ}$ and +37.5$^{\circ}$, respectively.
As shown in Figure~\ref{fig:scan}, the conformers are separated by a barrier of
almost 10 kJ/mol and differ in energy by 3.4 kJ/mol, the latter corresponding
to Boltzmann populations of 80\%:20\% at 296 K.

\begin{figure}
    \begin{minipage}[b]{0.22\textwidth}
        \begin{center}
            \includegraphics[width=\columnwidth]{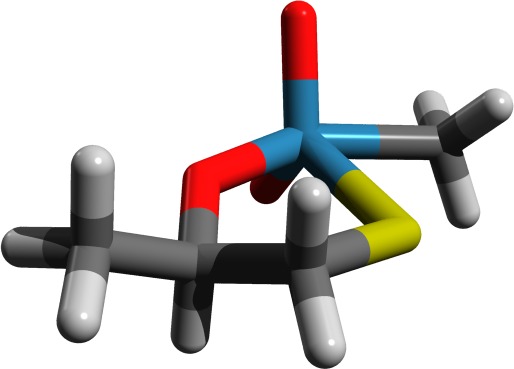} \\
            \textbf{3a-c1}
        \end{center}
    \end{minipage}%
    \begin{minipage}[b]{0.22\textwidth}
        \begin{center}
            \includegraphics[width=\columnwidth]{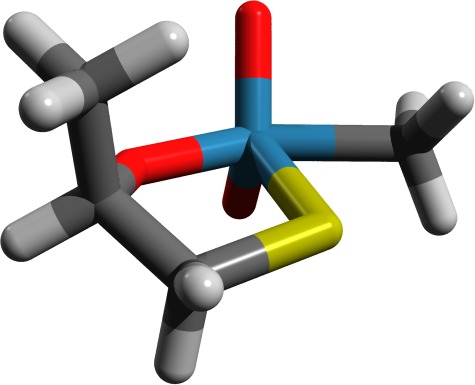} \\
            \textbf{3a-c2}
        \end{center}
    \end{minipage} \\
    \begin{minipage}[b]{0.22\textwidth}
        \begin{center}
            \includegraphics[width=\columnwidth]{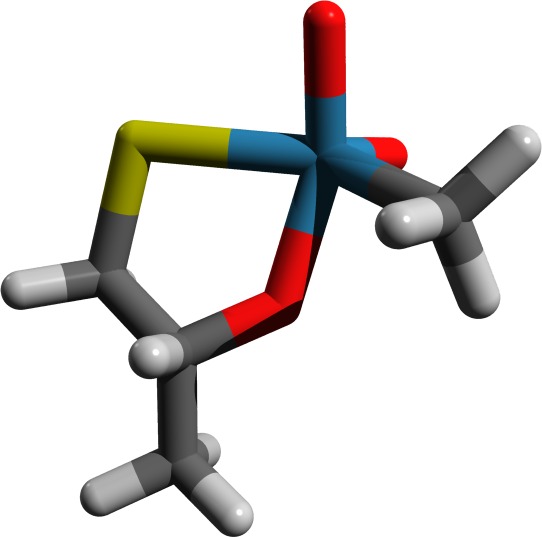} \\
            \textbf{3b-c1}
        \end{center}
    \end{minipage}%
    \begin{minipage}[b]{0.22\textwidth}
        \begin{center}
            \includegraphics[width=\columnwidth]{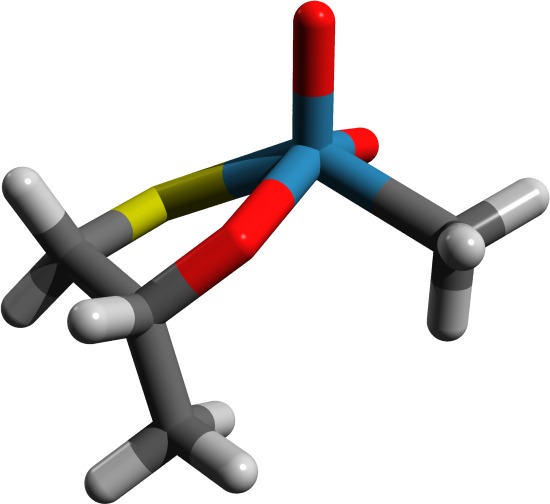} \\
            \textbf{3b-c2}
        \end{center}
    \end{minipage}
    \caption{
             \label{fig:conformers}
             B3LYP (def2-TZVPP basis) equilibrium structures of the theoretically
             studied conformers.
            }
\end{figure}

The conformers \textbf{3a-c1} and \textbf{3a-c2} are not the only possible
realizations of the ligation and we have extended the study
by optimizing the structures of corresponding \textbf{3b-c1} and \textbf{3b-c2}
conformers where the O--CHMe--CH$_2$--S ligand is connected to Re with
S--Re \emph{trans} to Re--Me.
The conformers \textbf{3b-c1} and \textbf{3b-c2} are separated by 5.4 kJ/mol,
corresponding to Boltzmann populations of 90\%:10\% at 296 K.  Conformer
\textbf{3b-c1} is, however, 18.4 kJ/mol higher in energy compared to \textbf{3a-c1}, and we
will in the following be able to confirm the predominance of form \textbf{3a-c1} from simulation of the spectra.  Before turning to the discussion of simulated spectra
we would first like to discuss the energetic separation between the four
studied conformers. To clarify whether the energetic preference for
\textbf{3a-c1} was sterically controlled by the Me group at the C2 atom or
electronically controlled according to the HSAB principle \cite{Pearson1963} we have optimized
structures for \textbf{3a'-c1}, \textbf{3a'-c2}, \textbf{3b'-c1}, and
\textbf{3b'-c2} conformers, where the Me group at the C2 atom has been replaced
by a H atom.  By doing this we find that the conformations \textbf{c1} and \textbf{c2} are
energetically almost equal for both \textbf{3a'} and \textbf{3b'} (separated by less than 0.02 kJ/mol,
but retaining the barrier heights)
while the difference between \textbf{3a'-c1} and
\textbf{3b'-c1} remains 18.0 kJ/mol, as compared to the difference of 18.4 kJ/mol between
\textbf{3a-c1} and \textbf{3b-c1}.  This indicates that the energy difference
between \textbf{c1} and \textbf{c2} is sterically controlled by the Me group attached to the C2
atom and can be tuned by bulkier ligands, while the preference for \textbf{3a}
over \textbf{3b} is electronically controlled by the O (HSAB harder) and S
(HSAB softer) atoms.\cite{Pearson1963} This theoretical study nicely illustrates the concept of
the \emph{trans}-effect in coordination chemistry.

\begin{figure}
    \includegraphics[width=\columnwidth]{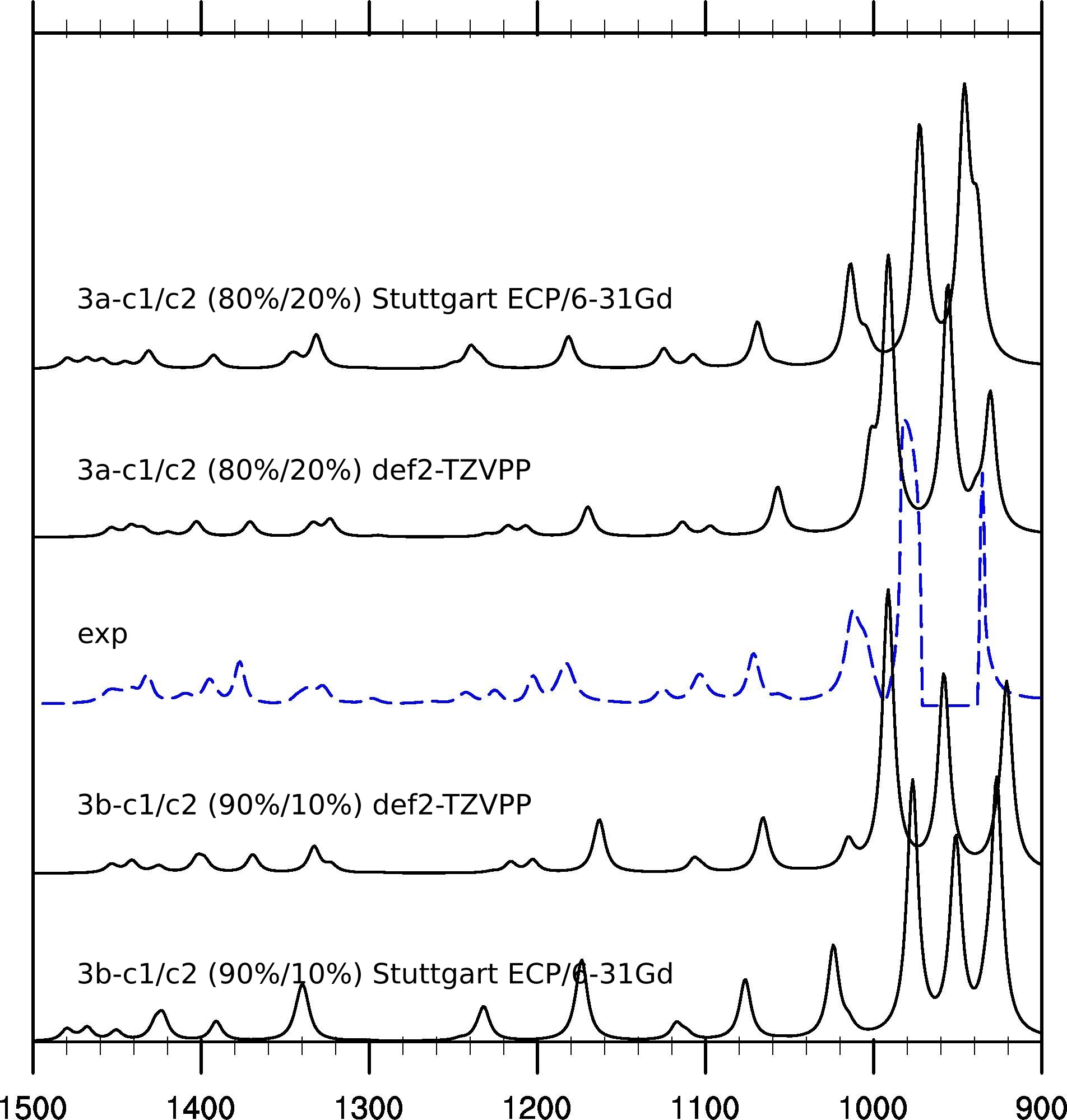}
    \caption{
             \label{fig:ir}
             Overlay of the simulated Boltzmann-averaged IR spectra for complex \textbf{3}
             (black solid lines, B3LYP, scaled),
             and the experimental spectrum (blue dashed line, 90 mg in 1 mL of CD$_2$Cl$_2$,
             100 $\mu$m cell, 4 cm$^{-1}$ resolution, 3000 scans).
            }
\end{figure}

\begin{figure}
    \includegraphics[width=\columnwidth]{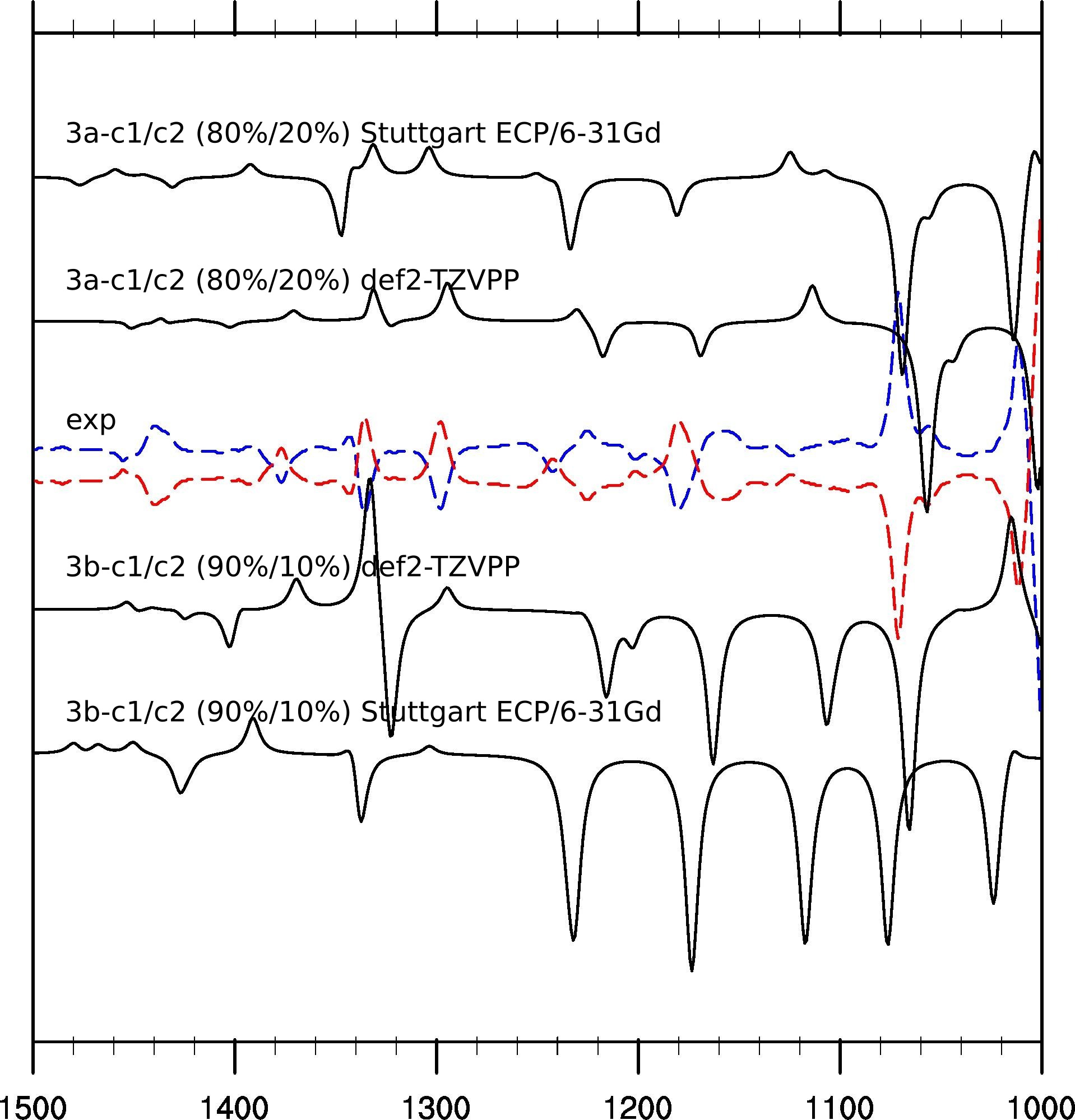}
    \caption{
             \label{fig:vcd}
             Overlay of the simulated Boltzmann-averaged VCD spectra for complex \emph{R}-\textbf{3}
             (black solid lines, B3LYP, scaled),
             and the experimental spectra (red and blue dashed lines correspond
             respectively to (+)-\textbf{3} and (-)-\textbf{3}, 90 mg in 1 mL of CD$_2$Cl$_2$,
             100 $\mu$m cell, 4 cm$^{-1}$ resolution, 3000 scans).
            }
\end{figure}

In order to verify the findings obtained using the relative energies and
to obtain more insight into the chiral environment around the metal
center, we next compared simulated IR and VCD spectra of complexes \textbf{3a} and
\textbf{3b} to the experimentally recorded spectra.  The experimentally recorded IR spectrum of complex \textbf{3a} in
CD$_2$Cl$_2$ is shown as a dashed line in
Figure~\ref{fig:ir}. The IR signature is
identical for both enantiomers (at this resolution).
The experimental VCD
spectra of \emph{R}-(+)-\textbf{3} and \emph{S}-(--)-\textbf{3} enantiomers in CD$_2$Cl$_2$ are shown in
Figure~\ref{fig:vcd} in red and blue color, respectively.
The experimental VCD spectra show a mirror-image relationship
and exhibit a strong
absorption band at 1012 cm$^{-1}$ (dissymmetry factor $\Delta \epsilon/\epsilon = 1.5 \times
10^{-4}$) which corresponds to
the oxorhenium symmetric bond stretching (the antisymmetric stretch, although
not accessible to our VCD spectrometer, is also visible on the experimental IR
spectrum of Figure~\ref{fig:ir} at $\sim$ 980 cm$^{-1}$).
The presence of
such a band indicates a chiral environment around the rhenium atom, although
the rhenium is not a stereogenic center in itself.

Comparing the observed IR and VCD spectra with the calculated spectra for the
Boltzmann-averaged \textbf{3a} and \textbf{3b} we note that the match for
conformer \textbf{3a} is not perfect, but it is better than for \textbf{3b},
which we expect based on energetic arguments.  We should also note that the
agreement between experimental and simulated spectra seems to be better for the
Stuttgart ECP/6-31Gd basis, already employed in Refs.
\citenum{Lassen2006} and \citenum{DeMontigny2009}, than for the def2-TZVPP basis,
although the latter is a more
flexible basis which we consider close to basis set limit at the DFT level.
This can probably be attributed to error cancellation effects.  Overall, the
VCD spectrum calculated for the \emph{R}- stereochemistry using the Stuttgart
ECP/6-31Gd basis set reproduces the experimental spectrum for the (+)
enantiomer well, except for the peak at 1160 cm\textsuperscript{-1}. The
discrepancy may be due to the flexibility of the five-membered rhenacycle in
solution that induces additional conformations. In conclusion this
conformational analysis and IR/VCD studies show that although the Re atom is
not strictly speaking a stereogenic center, the Re--O stretching bond at 1012
cm\textsuperscript{-1} has a strong well-defined chiroptical signature that is
a testimony of the chiral environment around the metal center. In our opinion
this makes the enantiopure chiral oxorhenium(VII) complex \textbf{3} a good
candidate for a PV measurement. For this reason, the PV vibrational shifts
associated with the Re--O stretching modes were predicted by relativistic DFT
calculations, as described in the following.

\section{Calculated PV shifts}\label{sec:pvres}

\begin{center}
    \begin{table*}[ht]
        \caption{\ Calculated reduced masses (in~amu),
		   vibrational transition frequencies (in~cm$^{-1}$) of the anti-symmetric and symmetric Re=O stretches for the conformers \textbf{3a-c1} and \textbf{3a-c2}, with corresponding 4-component DC
                   Hamiltonian PV fundamental transition frequency differences $\Delta \nu_{0\rightarrow 1}$ (in~Hz).
                   \label{tab:pnc_results}
                }
        \begin{tabular*}{\textwidth}
                    {
                     c r l r@{}l rr r@{}l
                    }
            \hline
            \hline
            \\
            Conformer
          & Mode
          & Method
          & \multicolumn{2}{c}{Red. mass}
          & Harmonic
          & Fundamental
          & \multicolumn{2}{c}{PV shift} \\
            \\
            \hline
            \\
            \textbf{3a-c1} & asym & HF    & 14.&2074 & 1106 & 1102 & --0.&211 \\ % ok
                           &      & B3LYP & 15.&7268 & 986  & 982  &   0.&078 \\ % ok
                           &      & PBE   & 15.&8522 & 951  & 944  &   0.&195 \\ % ok
            \\
                           & sym  & HF    &  8.&7326 & 1174 & 1165 &   0.&344 \\ % ok
                           &      & B3LYP & 12.&8095 & 1022 & 1010 &   0.&150 \\ % ok
                           &      & PBE   & 14.&8015 & 980  & 968  &   0.&036 \\ % ok
            \\
            \textbf{3a-c2} & asym & HF    & 15.&2424 & 1106 & 1102 &   0.&219 \\ % ok
                           &      & B3LYP & 15.&7515 & 985  & 981  & --0.&119 \\ % ok
                           &      & PBE   & 15.&8566 & 950  & 944  & --0.&271 \\ % ok
            \\
                           & sym  & HF    &  8.&4859 & 1175 & 1165 & --0.&056 \\ % ok
                           &      & B3LYP &  7.&2491 & 1021 & 1010 & --0.&170 \\ % ok
                           &      & PBE   & 13.&1114 & 979  & 966  & --0.&084 \\ % ok
            \\
            \hline
            \hline
        \end{tabular*}
    \end{table*}
\end{center}

\noindent Table~\ref{tab:pnc_results} summarizes the results of our computational study
of the selected two conformers \textbf{3a-c1} and \textbf{3a-c2} (see
Figure~\ref{fig:conformers}), where we have studied the symmetric and
anti-symmetric Re=O stretching mode frequencies using HF and two density
functional approximations, the hybrid B3LYP and the non-hybrid PBE
exchange-correlation functionals.  Comparing the 4-component DC Hamiltonian PV
fundamental transition frequency differences $\Delta \nu_{0 \rightarrow 1}$
(Table~\ref{tab:pnc_results}), we can first note that our calculations predict
PV vibrational frequency differences in the sub-Hz range, above the anticipated experimental
uncertainty in the frequency difference measurement.\cite{Darquie2010}
This is an encouraging result for our collaboration.  However, from
the theoretical point of view, the significant variation of the calculated PV
transition frequency differences is troubling, showing high
sensitivity to both the method applied (HF or density functional approximations) and to
structural changes.  Comparing HF with B3LYP and PBE we
see that the order of magnitude can change as well as the sign.
In principle we should rather trust the DFT calculations, since they incorporate electron
correlation, but the performance of DFT functionals with respect to PV shifts
for these heavy element compounds needs further study (in progress in our laboratory).
If we single out one method then a slight modification of the structural periphery seems to
have a major effect on the chiral environment around the heavy atom
(as exemplified by the sign change between the PV shifts for \textbf{3a-c1} and \textbf{3a-c2} in Table~\ref{tab:pnc_results}).
This makes parity violation \emph{chemically} interesting, as
detailed analysis of the underlying mechanism may provide a deeper understanding of
the electronic structure of chiral molecules. It should be emphasized that
conformational averaging, as carried out in the simulation of IR and VCD spectra in
section \ref{sec:vcd},
is not an issue in the ultra-high resolution experiment since it will resolve individual
lines of each conformer.

\section{Conclusion}
We have reported the synthesis and resolution, characterization, and
computational study of a new oxorhenium(VII) chiral compound. It is
designed for the molecular beam experiment, proposed in Ref. \citenum{Darquie2010},
dedicated to the observation of PV effects via high-resolution vibrational
spectroscopy. As an MTO derivative, this solid-state complex is expected
to sublime easily, a key aspect to allow gas phase studies. Measurements
and calculations show that the normal mode associated with the Re=O
stretch lies within the CO$_2$ laser operating range. Enantioselective
synthesis was developed for the preparation of enantioenriched complexes.
DFT calculations allowed the identification of four conformers
(\textbf{3a-c1}, \textbf{3a-c2}, \textbf{3b-c1} and \textbf{3b-c2}, see
Figure~\ref{fig:conformers}) as well as the rationalization of their relative stability.
We find that the stability \textbf{3a} with respect to \textbf{3b}
can be attributed to a \emph{trans} effect, whereas the smaller energy difference of \textbf{c1} with respect
to \textbf{c2} arises from a steric effect. The experimental VCD spectra are accordingly dominated by the two most stable conformers
\textbf{3a-c1} and \textbf{3a-c2}, as verified by their theoretical simulation which furthermore allowed determination of their
absolute configuration.

The VCD study has brought to light the chiral environment surrounding the rhenium atom, albeit not a stereogenic centre in itself.
Motivated by this observation, the vibrational transition frequency differences between the enantiomers
due to PV was calculated using Hartree--Fock (HF) and
DFT.
These results confirm the 0.1 Hz order of
magnitude for DFT PV vibrational frequency differences, as reported earlier
for oxorhenium complexes, which would be above the 0.01 Hz experimental
uncertainty expected in the frequency difference measurement.\cite{Darquie2010}
The present study puts emphasis on the sensitivity of the PV vibrational
frequency difference to the chemical environment around the rhenium
centre, as reported earlier by us.\cite{DeMontigny2010} This work represents an important
step towards the first experimental observation of PV in molecular
systems.

\section{Experimental details}\label{sec:exp}

\subsection{General}

Most experiments were performed using standard Schlenk techniques. Solvents
were freshly distilled under argon from sodium/benzophenone (THF) or from
phosphorus pentoxide (CH$_2$Cl$_2$). Starting materials were purchased from
ACBR (MTO) or from Aldrich. The synthesis of ligand \textbf{4} enantiomers was
performed by using a modified procedure.\cite{Fox2001} Column chromatography
purifications were performed in air over silica gel (Merck Geduran 60,
0.063--0.200 mm). \textsuperscript{1}H and \textsuperscript{13}C NMR spectra
were recorded on a Bruker AM300. \textsuperscript{13}C NMR spectra at 50.4 MHz
were recorded on a Bruker DPX 200. Chemical shifts were reported in parts per
million (ppm) relative to Si(CH$_3$)$_4$ as external standard and compared to
the literature. IR and VCD spectra were recorded on a Jasco FSV-6000
spectrometer.  Specific rotations (in deg
cm\textsuperscript{2}g\textsuperscript{-1}) were measured in a 10 cm
thermostated quartz cell on a Jasco P1010 polarimeter. Elemental analyses were
performed by the group CRMPO, University of Rennes 1.
CCDC reference number 910638 contains the crystallographic data for complex 3.
These data can be obtained free of charge at www.ccdc.cam.ac.uk/conts/retreving.html or
from the Cambridge Crystallographic Data Center, 12 union Road, Cambridge CB2 1EZ, UK;
Fax: (internat.) +44-1223-336-033; E-mail: deposit@ccdc.cam.ac.uk.

\textbf{
        \emph{S}-(--)-1-(tritylthio)propan-2-ol (\emph{S}-(--)-5).
       }
\emph{n}-BuLi (4.78 mmol, 2.5 M, 1.91 mL) was added dropwise to a
triphenylmethanethiol solution (4.67 mmol, 1.29 g) in 15 mL distilled THF
cooled at 0$^{\circ}$C, where a red-rose color persisted. \mbox{\emph{S}-(--)-propylene}
oxide (4.28 mmol, 0.3 mL) was then added dropwise at 0$^{\circ}$C with
a change in color to pale yellow. The reaction mixture was stirred over night,
quenched with 20\% AcOH in 20 mL methanol, diluted with water, and then
extracted with ethyl acetate. Purification over silica gel column
chromatography (pentane/ethyl acetate, 9:1) provided \emph{S}-(--)-\textbf{5}
with 90\% yield.  \textsuperscript{1}H NMR (300 MHz, CDCl$_3$) $\delta$
7.40--7.50 (6 H, m), 7.19--7.37 (9 H, m), 3.43 (1 H, sxt, $J$ = 6.2 Hz), 2.40
(1 H, d, $J$ = 1.1 Hz), 2.38 (1 H, s), 1.07 (3 H, d, $J$ = 6.1 Hz).
$[\alpha]_D^{23}$ = --54.4 ($C$ = 10\textsuperscript{-3} M, CH$_2$Cl$_2$)
ref.~\citenum{Fox2001}: $[\alpha]_D^{23}$ = --27.5 ($C$ = 3.35 $\times$ 10\textsuperscript{-3} M,
CHCl$_3$).  The same procedure was used for the preparation of the
\emph{R}-(+)-\textbf{5} enantiomer
$[\alpha]_D^{23}$ = +54.4 ($C$ = 10\textsuperscript{-3} M, CH$_2$Cl$_2$).

\textbf{
        \emph{S,S}-(+)-1,1'-disulfanediylbis(propan-2-ol) ((\emph{S,S})-(+)-6).
       }
\emph{S}-1-(tritylthio)propan-2-ol (\emph{S}-(--)-\textbf{5}) (300 mg, 0.897
mmol) dissolved in 50 mL DCM/Methanol (9:1) solution was added in portions (30
min) over Iodine solution, 1.1 g in 500 mL DCM/Methanol (9:1). The reaction
mixture was stirred at room temperature for 30 min, quenched with 10\% aqueous
Sodium thiosulfate and washed with brine. The organic layer was separated and
the aqueous layer was extracted with ethyl acetate twice, dried over MgSO$_4$,
and concentrated in vacuum to provide a dark brown precipitate which was
purified by silica gel chromatography (5\% ethanol/chloroform) to provide
(2\emph{S},2'\emph{S})-(+)-\textbf{6} as a yellow-brown oil (99\%).
\textsuperscript{1}H NMR (300 MHz, CDCl$_3$) $\delta$ 3.93--4.25 (1 H, m),
2.90 (1 H, dd, $J$ = 13.6, 3.4 Hz), 2.69 (1 H, dd, $J$ = 13.7, 8.5 Hz), 2.41 (1
H, br. s.), 1.31 (3 H, d, $J$ = 6.0 Hz).  \textsuperscript{13}C NMR (75 MHz,
CDCl$_3$) $\delta$ 65.9, 47.5, 22.0.  $[\alpha]_D^{23}$ = +232 ($C$ = 2.7 $\times$
10\textsuperscript{-3} M, CH$_2$Cl$_2$), ref.~\citenum{Fox2001}:
$[\alpha]_D^{23}$ = +199.3 ($C$ = 1.96 $\times$ 10\textsuperscript{-3} M,
CHCl$_3$).  The same procedure was used for the preparation of the
((\emph{R,R})-(--)-\textbf{6}) enantiomer $[\alpha]_D^{23}$ = --236 ($C$ = 2.7
$\times$ 10\textsuperscript{-3} M, CH$_2$Cl$_2$).

\textbf{
        \emph{S}-(+)-2-methyl-thiopropanol (\emph{S}-(+)-4).
       }
100 mg of (\emph{S,S})-1,1'-disulfanediylbis(propan-2-ol)
((\emph{S,S})-\textbf{6}) in THF was added dropwise to a stirring solution of
LAH (3 eq.) in THF at 0$^{\circ}$C, then stirred for 24 hrs at
50$^{\circ}$C. The reaction was quenched with AcOH and then extracted
with ether, dried over MgSO$_4$, and concentrated under vacuum to obtain
\emph{S}-(+)-\textbf{4} (42.2 mg, 84\%) as a colorless oil.
\textsuperscript{1}H NMR (300 MHz, CDCl$_3$) $\delta$ 3.77--3.86 (m, 1 H), 2.74
(dd, 1 H, $J$ = 14.1, 3.6 Hz), 2.5 (dd, 1 H, $J$ = 13.4 Hz, 7.4 Hz), 1.27 (d, 3
H, $J$ = 6.0 Hz).  $[\alpha]_D^{23}$ = +172 ($C$ = 5.10 $\times$
10\textsuperscript{-2} M, CH$_2$Cl$_2$).  The same procedure was used for the
preparation of the \emph{R}-(--)-\textbf{4} enantiomer.  $[\alpha]_D^{23}$ =
--166 ($C$ = 5.10 $\times$ 10\textsuperscript{-2} M, CH$_2$Cl$_2$).
\textsuperscript{13}C NMR (75 MHz, CDCl$_3$) $\delta$ 68.3, 32.8, 21.6. \cite{Fox2001}

\textbf{
        \emph{S}-(--)-3 complex.
       }
Methyltrioxorhenium (110 mg, 0.44 mmol) was dissolved in 10 mL distilled DCM,
\emph{S}-(+)-2-methyl-thiopropanol (0.44 mmol, 40.1 mg) was then
added, and the reaction mixture was stirred overnight under argon. Solvent was
removed under reduced pressure to obtain \emph{S}-(--)-\textbf{3} as a
yellow-orange precipitate.  \textsuperscript{1}H NMR (300 MHz, CDCl$_3$)
$\delta$ 5.30 (1 H, sxt, $J$ = 6.1 Hz), 3.84 (1 H, dd, $J$ = 11.3, 5.3 Hz),
3.46 (1 H, dd, $J$ = 11.7, 7.2 Hz), 2.50 (3 H, s), 1.51 (3 H, d, $J$ = 6.1 Hz).
\textsuperscript{13}C NMR (75 MHz, CDCl$_3$) $\delta$ 90.6, 46.2, 29.7, 20.8.
$[\alpha]_D^{23}$ = --17 ($C$ = 3.1 $\times$ 10\textsuperscript{-3} M,
CH$_2$Cl$_2$).  The same procedure was used for the preparation of the
\emph{R}-(+)-\textbf{3} enantiomer and for racemic \textbf{3}. $[\alpha]_D^{23}$ = +17 ($C$ = 3.1 $\times$
10\textsuperscript{-3} M, CH$_2$Cl$_2$).

\subsection{VCD measurements}

Samples of (+)-\textbf{3} and (--)-\textbf{3}, $\approx$ 9 mg/100
\ensuremath{\mu}L CD$_2$Cl$_2$, were placed in a 100-\ensuremath{\mu}m
path length cell with BaF$_2$ windows.  IR and VCD spectra were recorded on a
Jasco FVS-6000 VCD spectrometer with 3000 scans acquired and averaged at 4 cm\textsuperscript{-1}
resolution.  An overlay of the observed spectra for the two enantiomers, combined with the simulated Boltzmann-averaged VCD spectrum for complex \emph{R}-\textbf{3} in two different basis sets, is
presented in Figure~\ref{fig:vcd}.
The average VCD spectrum of the two enantiomers was used as the VCD baseline.

\section{Computational details}\label{sec:comp}

The optimized molecular structures and the corresponding IR and VCD spectra
were obtained using the Gaussian~09 package \cite{Gaussian09} employing the
default harmonic approximation \cite{Wilson_Decius_Cross:Molecular_Vibrations}
and using Kohn--Sham DFT together with the hybrid functional B3LYP\cite{Stephens1994, Becke1993} (employing
the VWN3\cite{Vosko1980} correlation functional).
In addition we have also employed HF and the PBE \cite{PBE} functional
to obtain the optimized structures and the harmonic force field for the subsequent
$E_\text{PV}$ calculations.
We have used
two different basis sets: Ahlrichs' def2-TZVPP \cite{def2-TZVPP, Re-ECP} (all
atoms) to obtain a result close to the basis set limit, and the less flexible
6-31Gd \cite{6-31Gd} (all atoms except Re) in combination with the
Stuttgart/Dresden ECP-60-MWB \cite{Re-ECP} (Re) for compatibility with a
previous VCD study of a similar compound.\cite{DeMontigny2009}  The vibrational
frequencies were scaled by 0.97 and the calculated intensities were converted
to Lorentzian bands with 4 cm$^{-1}$ half-width at half-maximum
for comparison to experiment.

In order to calculate the PV vibrational frequency shift we have performed
single-point energy and PV energy calculations using the DIRAC12 program
\cite{DIRAC12} following the
normal mode coordinates
taken from the corresponding
Gaussian~09 calculation. The single-point energies, now based on
the 4-component relativistic Dirac-Coulomb (DC) Hamiltonian,
calculated along the effective core potential normal mode provided the
anharmonic potential for the numerical solution of the vibrational wave
functions by the Numerov--Cooley algorithm.\cite{Numerov1933, Cooley1961}

The PV vibrational frequency shift $v_{m \rightarrow n}$ for the
$m \rightarrow n$ vibrational transition (in our case $0 \rightarrow 1$) is obtained using
\begin{equation}
    h\Delta \nu_{m \rightarrow n} = 2\left( \langle n \vert E_\text{PV} (q_R) \vert n \rangle
                        -  \langle m \vert E_\text{PV} (q_R) \vert m \rangle\right) ,
\end{equation}
where $\vert n \rangle$ and $\vert m \rangle$ are the corresponding Numerov--Cooley
vibrational wave functions. $E_\text{PV} (q_R)$ is the PV energy along
the normal mode coordinate $q$ of the $R$ enantiomer, evaluated as an
expectation value of the operator
\begin{equation}
    \hat{H}_\text{PV}     = \sum_K \hat{H}_{\text{PV},K};
    \qquad
    \hat{H}_{\text{PV},K} = \frac{G_\text{F}}{2\sqrt{2}} Q_{\text{w},K} \sum_{i} \gamma^5 \rho_K (\mathbf{r}_{iK}),
\end{equation}
in which appears the weak nuclear charge $Q_{\text{w},K} = Z_K (1 - 4 \sin^2
\theta_\text{W}) - N_K$ with $Z_K$ and $N_K$ representing the number of protons
and neutrons in nucleus $K$, respectively, and $\sin^2 \theta_\text{W} =
0.2319$ is the employed Weinberg parameter.  The normalized nuclear charge
densities $\rho_K$, in this work represented by Gaussian distributions,
\cite{Visscher1997} restrict integration over electron
coordinates $\mathbf{r}_i$ to nuclear regions, thus providing a natural
partitioning of the operator in atomic contributions $\hat{H}_{\text{PV},K}$.
Finally, $\gamma^5$ is one of the Dirac matrices, Eq.~(\ref{equation:gamma5}),
with $1_{2\times2}$ and $0_{2\times2}$ being the $2\times 2$ identity and null matrices, respectively
\begin{equation} \label{equation:gamma5}
    \gamma^5 = \begin{bmatrix}
                   0_{2\times 2} & 1_{2\times 2} \\
                   1_{2\times 2} & 0_{2\times 2}
               \end{bmatrix}.
\end{equation}
The Fermi coupling constant $G_\text{F} = 2.22254 \times 10^{-14} E_\text{h}
a_0^3$ demonstrates the minuteness of the PV effect.

We have calculated PV energies at the HF level as well as using the density functionals
B3LYP\cite{Stephens1994, Becke1993} and
PBE,\cite{PBE} all in a 4-component relativistic framework.
For the DC single-point calculations we have employed
uncontracted cc-pVDZ basis sets \cite{cc-pVDZ} for all atoms except Re,
for which we have used the basis sets employed in Ref.~\citenum{DeMontigny2010}.
The small component basis sets were generated by restricted kinetic balance
imposed in the canonical orthonormalization step.
\cite{Visscher2000} The two-electron Coulomb integrals
(SS$\vert$SS), involving only the small components, were neglected in all
calculations and the energy corrected by a simple point-charge model.
\cite{Visscher1997b}

\section{Acknowledgment}

We thank the Minist{\`e}re de l'Education Nationale, de la Recherche et de la
Technologie, and the Centre National de la Recherche Scientifique (CNRS). This
work is part of the project NCPCHEM 2010 BLAN 724 3 funded by the Agence
Nationale de la Recherche (ANR, France).  This work has received support from
the high performance computing centre ``Calcul en Midi-Pyr\'en\'ees'' (CALMIP)
through a grant of computer time.


\begin{thebibliography}{99}

\bibitem{Francotte2006}
``Chirality in Drug Research'',
E. Francotte, W. Lindner (eds.), Wiley-VCH, 2006.

\bibitem{Amabilino2009}
``Chirality at the Nanoscale: Nanoparticles, Surfaces, Materials and more'',
D. Amabilino (ed.), Wiley-VCH, 2009.

\bibitem{Amabilino2009b}
Special issue on chirality at the nanoscale:
D. Amabilino (Guest Editor),
\emph{Chem. Soc. Rev}. 2009, \textbf{38}, 659.

\bibitem{Mikami2007}
``New Frontiers in Asymmetric Catalysis'',
K. Mikami, M. Lautens (eds.), Wiley-VCH, 2007.

\bibitem{Viedma2011}
C. Viedma, P. Cintas
\emph{Isr. J. Chem.}, 2011, \textbf{51}, 997 and references therein.

\bibitem{Wagniere2007}
``On Chirality and the Universal Asymmetry: Reflections on Image and Mirror Image'',
G. H. Wagni\`ere (ed.), Wiley-VCH, 2007.

\bibitem{Quack2002}
M. Quack,
\emph{Angew. Chem. Int. Ed.}, 2002, \textbf{41}, 4618.

\bibitem{Avalos2000}
M. Avalos, R. Babiano, P. Cintas, J. L. Jimenez, J. C. Palacios,
\emph{Tetrahedron Asymmetry}, 2000, \textbf{11}, 2845.

\bibitem{Crassous2003}
J. Crassous, F. Monier, J.-P. Dutasta, M. Ziskind, C. Daussy, C. Grain and C. Chardonnet,
\emph{ChemPhysChem}, 2003, \textbf{4}, 541.

\bibitem{Crassous2005}
J. Crassous, C. Chardonnet, T. Saue and P. Schwerdtfeger,
\emph{Org. Biomol. Chem.}, 2005, \textbf{3}, 2218.

\bibitem{Chardonnet2006}
C. Chardonnet, C. Daussy, O. Lopez, A. Amy-Klein, In:
Maroulis G, Simos T (eds.)
``Trends and Perspectives in Modern Computational Science.'' Vol. 6, Lecture Series
on Computer and Computational Sciences, 2006. 324--331.

\bibitem{Darquie2010}
B. Darqui\'e, C. Stoeffler, A. Shelkovnikov, C. Daussy, A. Amy-Klein,
C. Chardonnet, S. Zrig, L. Guy, J. Crassous, P. Soulard, P. Asselin, T. R. Huet,
P. Schwerdtfeger, R. Bast, T. Saue,
\emph{Chirality}, 2010, \textbf{22}, 870.

\bibitem{Stoeffler2011}
C. Stoeffler, B. Darqui\'e, A. Shelkovnikov, C. Daussy, A.
Amy-Klein, C. Chardonnet, L. Guy, J. Crassous, T. R. Huet, P. Soulard, P. Asselin,
\emph{Phys. Chem. Chem. Phys.}, 2011, \textbf{13}, 854.

\bibitem{Laerdahl2000}
J. K. L{\ae}rdahl, P. Schwerdtfeger, H. M. Quiney,
\emph{Phys. Rev. Lett.}, 2000, \textbf{84}, 3811.

\bibitem{Schwerdtfeger2004}
P. Schwerdtfeger and R. Bast,
\emph{J. Am. Chem. Soc.}, 2004, \textbf{126}, 1652.

\bibitem{Schwerdtfeger2003}
P. Schwerdtfeger, J. Gierlich, T. Bollwein,
\emph{Angew. Chem. Int. Ed.}, 2003, \textbf{42}, 1293.

\bibitem{Faglioni2003}
F. Faglioni, P. Lazzeretti,
\emph{Phys. Rev. A}, 2003, \textbf{67}, 032101.

\bibitem{Bast2003}
R. Bast, P. Schwerdtfeger,
\emph{Phys. Rev. Lett.}, 2003, \textbf{91}, 023001.

\bibitem{Figgen2010}
D. Figgen, A. Koers, P. Schwerdtfeger,
\emph{Angew. Chem. Int. Ed.}, 2010, \textbf{49}, 2941.

\bibitem{Bock2006}
F. Bock, F. Fischer, W. A. Schenk,
\emph{J. Am. Chem. Soc.}, 2006, \textbf{128}, 68.

\bibitem{Merrifield1982}
J. H. Merrifield, C. E. Strouse, J. A. Gladysz,
\emph{Organometallics}, 1982, \textbf{1}, 1204.

\bibitem{Buhro1983}
W. E. Buhro, A. Wong, J. H. Merrifield, G.-Y. Lin, A. C. Constable, J. A. Gladysz,
\emph{Organometallics}, 1983, \textbf{2}, 1852.

\bibitem{Lassen2006}
P. R. Lassen, L. Guy, I. Karame, T. Roisnel, N. Vanthuyne, C. Roussel, X. Cao, R. Lombardi, J. Crassous,
T. B. Freedman, L. A. Nafie,
\emph{Inorg. Chem.}, 2006, \textbf{45}, 10230.

\bibitem{DeMontigny2009}
F. De Montigny, L. Guy, G. Pilet, N. Vanthuyne, C. Roussel, R. Lombardi, T. B.
Freedman, L. A. Nafie, J. Crassous,
\emph{Chem. Comm.}, 2009, 4841.

\bibitem{DeMontigny2010}
F. De Montigny, R. Bast, A. Severo Pereira Gomes, G. Pilet, N. Vanthuyne, C. Roussel,
L. Guy, P. Schwerdtfeger, T. Saue, J. Crassous,
\emph{Phys. Chem. Chem. Phys}. 2010, \textbf{12}, 8792.

\bibitem{Faller2000}
J. W. Faller, A. R. Lavoie,
\emph{Organometallics}, 2000, \textbf{19}, 3957.

\bibitem{Rybak2003}
W. K. Rybak, A. Skarzynska, T. Glowiak,
\emph{Angew. Chem. Int. Ed.}, 2003, \textbf{42}, 1725.

\bibitem{Tazacs2012}
E. Tazacs, A. Escande, N. Vanthuyne, C. Roussel, C. Lescop, E. Guinard,
C. Latouche, A. Boucekkine, J. Crassous, R. R\'eau, M. Hissler,
\emph{Chem. Comm.}, 2012, \textbf{48}, 6705.

\bibitem{Jain2008}
K. R. Jain, W. A. Herrmann, F. E. Kuhn,
\emph{Coord. Chem. Rev.}, 2008, \textbf{252}, 556.

\bibitem{Owen2000}
G. S. Owen, J. Arias, M. M. Abu-Omar,
\emph{Catalysis Today}, 2000, \textbf{55}, 317.

\bibitem{Harding2000}
R. L. Harding, T. D. H. Bugg,
\emph{Tetrahedron Lett.}, 2000, \textbf{41}, 2729.

\bibitem{Fox2001}
B. M. Fox, J. A. Vroman, P. E. Fanwick, M. Cushman,
\emph{J. Med. Chem.}, 2001, \textbf{44}, 3915.

\bibitem{Dixon2002}
J. Dixon, J. H. Espenson,
\emph{Inorg. Chem.}, 2002, \textbf{41}, 4727.

\bibitem{Espenson2002}
J. H. Espenson, X. Shan, Y. Wang, R. Huang, D. W. Lahti, J. Dixon, G. Lente, A. Ellern, I. A. Guzei,
\emph{Inorg. Chem.}, 2002, \textbf{41}, 2583.

\bibitem{Freedman2003}
T. B. Freedman, X. Cao, R. K. Dukor, L. A. Nafie,
\emph{Chirality}, 2003, \textbf{15}, 743.

\bibitem{Stephens2000}
P. J. Stephens, F. J. Devlin,
\emph{Chirality}, 2000, \textbf{12}, 172.

\bibitem{Polavarapu2004}
P. L. Polavarapu, J. He,
\emph{Anal. Chem.}, 2004, \textbf{76}, 61A.

\bibitem{Chamayou}
A. -C. Chamayou, S. Luedeke, V. Brecht, T. B. Freedman, L. A. Nafie, C. Janiak,
\emph{Inorg. Chem.}, 2011, \textbf{50}, 11363.

\bibitem{def2-TZVPP}
    F. Weigend, R. Ahlrichs,
    \emph{Phys. Chem. Chem. Phys.}, 2005, \textbf{7}, 3297.

\bibitem{Pearson1963}
R. G. Pearson,
\emph{J. Am. Chem. Soc.}, 1963, \textbf{85}, 3533.

\bibitem{Gaussian09}
    Gaussian 09, Revision A.1, M. J. Frisch, G. W. Trucks, H. B. Schlegel, G. E.
    Scuseria, M. A. Robb, J. R. Cheeseman, G. Scalmani, V. Barone, B. Mennucci, G.
    A. Petersson, H. Nakatsuji, M. Caricato, X. Li, H. P. Hratchian, A. F.
    Izmaylov, J. Bloino, G. Zheng, J. L. Sonnenberg, M. Hada, M. Ehara, K. Toyota,
    R. Fukuda, J. Hasegawa, M. Ishida, T. Nakajima, Y. Honda, O. Kitao, H. Nakai,
    T. Vreven, J. A. Montgomery, Jr., J. E. Peralta, F. Ogliaro, M. Bearpark, J. J.
    Heyd, E. Brothers, K. N. Kudin, V. N. Staroverov, R. Kobayashi, J. Normand, K.
    Raghavachari, A. Rendell, J. C. Burant, S. S. Iyengar, J. Tomasi, M. Cossi, N.
    Rega, J. M. Millam, M. Klene, J. E. Knox, J. B. Cross, V. Bakken, C. Adamo, J.
    Jaramillo, R. Gomperts, R. E. Stratmann, O. Yazyev, A. J. Austin, R. Cammi, C.
    Pomelli, J. W. Ochterski, R. L. Martin, K. Morokuma, V. G. Zakrzewski, G. A.
    Voth, P. Salvador, J. J. Dannenberg, S. Dapprich, A. D. Daniels, {\"O} Farkas, J.
    B. Foresman, J. V. Ortiz, J. Cioslowski, and D. J. Fox, Gaussian, Inc.,
    Wallingford CT, 2009.

\bibitem{Wilson_Decius_Cross:Molecular_Vibrations}
    E.~B. Wilson, J.~C. Decius and P.~C. Cross,
    \emph{Molecular Vibrations: The Theory of Infrared and Raman Vibrational Spectra},
    Dover Publications, Inc., New York, 1995.

\bibitem{Stephens1994}
    P.~J. Stephens, F.~J. Devlin, C.~F. Chabalowski and M.~J. Frisch,
    \emph{J. Phys. Chem}, 1994, \textbf{98}, 11623.

\bibitem{Becke1993}
    A.~D. Becke,
    \emph{J. Chem. Phys.}, 1993, \textbf{98}, 5648.

\bibitem{Vosko1980}
    S. J. Vosko, L. Wilk, M. Nusair,
    \emph{Can. J. Phys.}, 1980, \textbf{58}, 1200.

\bibitem{PBE}
    J. P. Perdew, K. Burke, M. Ernzerhof,
    \emph{Phys. Rev. Lett.}, 1996, \textbf{77}, 3865.

\bibitem{Re-ECP}
    D. Andrae, U. Haeussermann, M. Dolg, H. Stoll, H. Preuss,
    \emph{Theor. Chim. Acta}, 1990, \textbf{77}, 123.

\bibitem{6-31Gd}
    M. J. Frisch, J. A. Pople, J. S. Binkley,
    \emph{J. Chem. Phys.} 1984, \textbf{80}, 3265;
    M. M. Francl, W. J. Petro, W. J. Hehre, J. S. Binkley, M. S. Gordon, D. J. DeFrees and J. A. Pople,
    \emph{J. Chem. Phys.} 1982, \textbf{77}, 3654.

\bibitem{DIRAC12}
    DIRAC, a relativistic ab initio electronic structure program,
    Release DIRAC12 (2012),
    written by H.~J.~{\relax Aa}.~Jensen, R.~Bast, T.~Saue, and L.~Visscher,
    with contributions from V.~Bakken, K.~G.~Dyall, S.~Dubillard,
    U.~Ekstr{\"o}m, E.~Eliav, T.~Enevoldsen, T.~Fleig, O.~Fossgaard,
    A.~S.~P.~Gomes, T.~Helgaker, J.~K.~L{\ae}rdahl, Y.~S.~Lee, J.~Henriksson,
    M.~Ilia{\v{s}}, Ch.~R.~Jacob, S.~Knecht, S.~Komorovsk{\'y}, O.~Kullie, C.~V.~Larsen, H.~S.~Nataraj,
    P.~Norman, G.~Olejniczak, J.~Olsen, Y.~C.~Park, J.~K.~Pedersen, M.~Pernpointner,
    K.~Ruud, P.~Sa{\l}ek, B.~Schimmelpfennig, J.~Sikkema, A.~J.~Thorvaldsen,
    J.~Thyssen, J.~van~Stralen, S.~Villaume, O.~Visser, T.~Winther,
    and S.~Yamamoto
    (see http://www.diracprogram.org).

\bibitem{Numerov1933}
    B.~Numerov,
    \emph{Publ. Obs. Central Astrophys. Russ.}, 1933, \textbf{2}, 188.

\bibitem{Cooley1961}
     J.~W. Cooley,
     \emph{Math. Comput.}, 1961, \textbf{15}, 363.

\bibitem{Visscher1997}
    L.~Visscher, K.~G. Dyall,
    \emph{At. Data Nucl. Data Tables}, 1997, \textbf{67}, 207.

\bibitem{cc-pVDZ}
    T. H. Dunning Jr.,
    \emph{J. Chem. Phys.}, 1989, \textbf{90}, 1007;
    D. E. Woon, T. H. Dunning Jr.,
    \emph{J. Chem. Phys.}, 1993, \textbf{98}, 1358.

\bibitem{Visscher2000}
    L.~Visscher, T.~Saue,
    \emph{J. Chem. Phys.}, 2000, \textbf{113}, 3996.

\bibitem{Visscher1997b}
    L.~Visscher,
    \emph{Theor. Chem. Acc.}, 1997, \textbf{89}, 68.


\end{thebibliography}
\end{document}